\documentclass[pra,twocolumn,showpacs,preprintnumbers,amsmath,amssymb,showkeys]{revtex4}

\usepackage{epsf}
\usepackage{amssymb}

\newcommand{\ket}[1]{\mathop{\left| #1 \right\rangle}\nolimits}
\newcommand{\bra}[1]{\mathop{\left\langle #1 \right|}\nolimits}
\newcommand{\braket}[2]{\mathop{\left\langle #1 \left| #2 \right.
\right\rangle}\nolimits}

\begin{document}

\title{Physical implementation of entangling quantum measurements}

\author{Boris A. Grishanin}
 \email{grishan@comsim1.phys.msu.ru}
\author{Victor N. Zadkov}
 \email{zadkov@comsim1.phys.msu.ru}
\affiliation{International Laser Center and Department of Physics\\
M.\ V.\ Lomonosov Moscow State University, 119899 Moscow, Russia}

\date{October 22, 2004}

\begin{abstract}
We clarify the microscopic structure of the entangling quantum measurement
superoperators and examine their possible physical realization in a simple
three-qubit model, which implements the entangling quantum measurement with an
arbitrary degree of entanglement.
\end{abstract}



\maketitle

\section{\label{section:introduction}Introduction}

In quantum information theory, generalized description of most important quantum
transformations, which extend the class of unitary transformations lying in the
foundations of quantum theory of dynamically closed quantum systems \cite{a}, plays
very important role \cite{b}. Particulary, the resulting transformation in a system
describing only the measured object to which we apply the standard quantum
measurement can be written in a form of so called projective measurement
superoperator:
\begin{equation}\label{MP}
{\cal M}_P=\sum \ket k\bra k\odot\ket k\bra k,
\end{equation}

\noindent where the $k$-terms of the sum describe the normalized
positive superoperator measure (PSM), which is represented here by
the orthogonal projection superoperators of the form
\[
{\cal E}_k=\ket k\bra
k\odot\ket k\bra k,\quad \sum{\rm Tr}\,{\cal E}_k={\rm Tr}\,\odot.
\]

\noindent Respectively, $\sum{\cal E}_k^+\hat I=\sum\ket k\bra k=\hat I$ (see, for
instance, \cite{c}). The substitution symbol $\odot$ is to be substituted by a
transformed operator, which is simply the density matrices in our case \cite{d,e};
index $k$ enumerates the eigen vectors of the measured physical variable, which is
described by the operator $\hat A=\sum\lambda_k\ket k\bra k$ in the Hilbert space
$H_A$ of the measured object. The generalized measurement, which is carried out in
the extended space $H_A\otimes H_a$ of the initial and auxiliary systems, is
described by the PSM of the general form ${\cal E}_k=\hat F_k^{}\odot \hat F_k^+$ in
the linear space of operators in $H_A$. The corresponding classical probabilistic
measure on the spectrum of physically possible values $\lambda_k$ of the operator
$\hat A$ is described by the linear functional $\hat\rho_A\to P(k)={\rm Tr}\,\hat
E_k\hat\rho_A$, which is determined by a non-orthogonal expansion of the unit
operator \cite{f}
\[
\hat E_k=\hat F_k^+\hat F_k^{},\quad \sum\hat E_k=\hat I,
\]

\noindent which is the positive operator valued measure (POVM) \cite{g}.

Due to the progress in quantum state engineering made over last decades \cite{h},
the commonly accepted concept of quantum measurement as a projective transformation
has been essentially revised. It includes now various types of measurements, e.g.,
the measurement that provides the measured information in a form of quantum
entanglement between the apparatus and the measured object. By contrast with the
classical theory, the equality $A\equiv B$, which means the coincidence of the
physical values $A$ and $B$ for all their possible values $\lambda$, can be realized
now differently. This equality does not prevent arbitrary relations between the
phases $\varphi_\lambda$ corresponding to the eigen wave-functions $\Psi_\lambda$
implementing the equality $A=B=\lambda$. Therefore, the standard quantum measurement
implies complete absence of the phase correlations, whereas the completely coherent
measurement implies the defined set of phases.

The respective most general abridged notation for the ideal quantum measurement
transformation in the object--ap\-pa\-ra\-tus system is given by the
\emph{entangling} quantum measurement superoperator \cite{e}. This entangling
quantum measurement can be considered as a combination of the completely coherent
measurement, which provides the measurement results in a form of quantum
entanglement between the apparatus and the object, and additional transformation
dephasing the states of the apparatus
$${\cal D}=\sum_{ij}R_{ij}\ket i\bra i\odot\ket j\bra j
$$

\noindent with the positive entanglement matrix $R\ge0$ with the
diagonal elements $R_{ii}\equiv1$. The entangling quantum
measurement is an intermediate transformation between the identity
superoperator transformation ${\cal I}=\sum_{ij}\ket i\bra
i\odot\ket j\bra j$, which corresponds to the case of
$R_{ij}\equiv1$, and the projective measurement transformation
(\ref{MP}), which corresponds to the diagonal matrix
$R_{ij}=\delta_{ij}$.

Definition of quantum measurement considered in Ref.~\cite{e} is based on the
natural interpretation of the quantum measurement as the transformation, which is
invariant with regard to the initial state of the apparatus. However, in a wide
range of experimental situations \cite{h,i,j,k} the quantum measurement
transformations are applied to a bipartite system when the initial state of one of
the subsystems is explicitly known (it can be, for example, the ground or specially
prepared quantum state of an atom or non-excited resonator mode). Both cases can be
described with the superoperators of a specialized type, which instead of the
complete mapping (object+apparatus) $\to$ (object+apparatus) define the mapping
(object) $\to$ (object+apparatus). It is worth to note here that for a potentially
capable experimental realization of the measurement transformations considering
appropriate mathematical representation of a specific physical situation is of prime
importance.

In this work, we elucidate possibility of physical implementation
of an entangling measurement. The general theory is illustrated on
example of two-level models, which describe in an idealized form
some features of quantum transformations that are typical, for
instance, for the experiments with trapped atoms.

In Sec.\ \ref{section:expanded} we give precise mathematical
definitions of the extended superoperators and discuss how they
can be applied to the various types of the ideal quantum
measurement. Also, we give physical interpretation of the coherent
information, which is bound to the entangling quantum measurement.
Unitary implementations of the extended
superoperators in connection with the experimental specificity of
physical implementations of non-reversal transformations are
considered in Sec.\ \ref{section:unitary}. In Sec.\
\ref{section:matrix} we consider specific matrix representations
in application to the extended superoperators technique. In Sec.\
\ref{section:2D} we specify a unitary realization of the
entangling measurement in a simple three-qubit model, which
implements the entangling quantum measurement with an arbitrary
degree of entanglement.

\section{\label{section:expanded}Mathematical definitions of extended superoperators}

In order to clarify physical implementation of the measurement
transformations, its mathematical representation has to have a
clear and simple form. Extended superoperators perfectly fit this
purpose and their definitions are considered below in detail.

Let us consider a superoperator transformation ${\cal S}$ in a bipartite system
$A$+$B$, which we apply to the density matrix $\hat\rho_{AB}^{}=
\hat\rho_A^{}\otimes\hat\rho_B^0$ of this system in the Hilbert space $H_A\otimes
H_B$, where $\hat\rho_B^0$ is an arbitrary chosen fixed state. Then, the result of
this superoperator transformation is simply a map $\mathbb{C}(H_A)\to\mathbb
{C}(H_A\otimes H_B)$ of the operators algebra in $H_A$ onto the corresponding
algebra in $H_A\otimes H_B$ and can be written in a symbolic representation as
\begin{equation}\label{expand}
{\cal E}={\cal S}(\odot\otimes\hat\rho_B^0),
\end{equation}

\noindent where the substitution symbol $\odot$ should be
substituted by single transformed operator $\hat\rho_A$. By
contrast with ${\cal S}$, the \emph{extended superoperator} ${\cal
E}$ has an ``extended" in comparison with the input
$\hat\rho_A^{}$ space with the elements $\hat\rho_{AB}^{}={\cal
E}\hat\rho_A^{}$.

If the result of the superoperator transformation does not depend
on $\hat\rho_B^0$, i.e., ${\cal E}={\cal E}_0\equiv {\cal
S}(\odot\otimes\hat\rho_B^{})$ for all $\hat\rho_B$, we have
another special case, when the superoperator transformation ${\cal
S}$ can be described entirely with the extended superoperator
(\ref{expand}). The corresponding structure of such invariant
superoperator has the form:
\begin{equation}\label{inv}
{\cal S}=\sum_{ij}\Bigl({\cal S}^A_{ij}\,\odot_A\Bigr)\otimes\Bigl(\ket i
\bra j{\rm Tr}\,\odot_B\Bigr),
\end{equation}

\noindent where the trace operation makes the result independent of an initial
state of the system $B$.

The extended superoperator (\ref{expand}) may be treated as a ``hybrid"
superoperator transformation over the variables of the system $A$ and the density
matrix operator over the variables of the system $B$. Respectively, tracing the
extended superoperator over the variables of the system $B$ results in a regular
superoperator ${\cal S}_A={\rm Tr}_B^{}{\cal E}$, which maps algebra $\mathbb
{C}(H_A)$ onto itself. The relation ${\cal S}_A\to{\cal E}$ can be considered as an
extension of the value area of the superoperator, which is related to the concrete
definition of the respective physical transformation in an open system in the
symbolic representation form (\ref{expand}).

Apparently, the extended superoperator (\ref{expand}) has the same
specificity for all superoperators properties---the complete
positivity and normalization. In case of $d$-dimensional Hilbert
spaces $H_A$, $H_B$, the extended superoperator can be represented
in the matrix representation by the rectangular matrices of
$d^4\times d^2$-dimension, whereas a regular superoperator ${\cal
S}$ is described by rectangular matrices of $d^4\times
d^4$-dimension. In a specific case of two qubits, these are
$16\times4$ and $16\times16$ matrices, respectively. Keeping this
in mind, one can essentially reduce complexity of the respective
calculations performing them in terms of the extended
superoperators, when it is possible.

With the help of an orthogonal basis $\ket k$ in $H_A$ the
extended superoperator (\ref{expand}) has the following, as one
can easily see, most generalized form:
\begin{equation}\label{gen}
{\cal E}=\sum\hat s_{kl}\bra k\odot\ket l,
\end{equation}

\noindent where $\hat s_{kl}$ is the set of operators in $H_{A}\otimes H_B$,
which satisfy the above mentioned complete positivity and normalization conditions.

From the properties of the extended superoperators it is also follows that more than
one regular superoperator can correspond to the extended one. Also, possibility of
physical implementation of the extended superoperator ${\cal E}$, which satisfies
the complete positivity condition, readily follows from the general criterion of
physical implementation of a regular superoperator \cite{b} and it is enough to have
only the existence proof of a complete positive superoperator ${\cal S}$ and density
matrix $\rho_B^0$, related to ${\cal E}$ according to the Eq.\ (\ref{expand}).

Let us consider the entangling measurement superoperator:
\begin{equation}\label{M}
{\cal M}=\sum_{ijm} R_{ij}\ket i\ket i\bra j\bra j\bra m\bra i\odot\ket
j\ket m
\end{equation}

\noindent with the entanglement matrix $(R_{ij})\ge0,R_{ii}\equiv1$ \cite{e}, which
is a particular case of the invariant superoperator (\ref{inv}). The resulted state
after its action does not depend on an initial state of the system $B$ and with the
help of (\ref{gen}) the corresponding extended entangling measurement superoperator
has the form:
\begin{equation}\label{EM}
{\cal E}_M=\sum_{ij}\bigl(R_{ij}\ket i\ket i\bra j\bra j\bigr)\bra
i\odot\ket j.
\end{equation}

\noindent Here, the resulted state $\hat\rho_{AB}$ is represented
only via the cloned basis states $\ket i\ket i$, which means that
the quantum measurement was an ideal one. Also, a fact that
$R_{ij}\ne1$ at $i\ne j$ is an evidence that the measurement is an
incoherent one. Even in the case of complete coherency,
$R_{ij}\equiv1$, when the entangling superoperator (\ref{M})
describes \emph{cloning transformation} of the basis states,
\begin{equation}\label{C}
{\cal C}=\sum_{ijm}\ket i\ket i\bra j\bra j\bra m\bra i\odot\ket j\ket m,
\end{equation}

\noindent it is \emph{non-reversal} because information of an initial state of the
apparatus $\hat\rho_B$ is completely ignored.

The extended superoperator for the entangling measurement can be
additionally extended in a way to clarify the quantum nature of
the entanglement matrix. This can be readily done by introducing
additional internal degrees of freedom in $H_D$ space that are
responsible for the dephasing effects, i.e., in the form of the
extended superoperator $A\to (A{+}B{+}D)$ of the form:
\begin{equation}\label{EABD}
{\cal E}_M=\sum_{ij} \ket i\ket i\ket{\ket i}\bra{\bra j}\bra j\bra j\bra
i\odot\ket j,
\end{equation}

\noindent where internal degrees of freedom are in the double brackets and, generally,
are non-orthogonal and are described by the scalar product
$\left<\braket{i}{j}\right>=R_{ij}$, which ensures coincidence with Eq.\ (\ref{EM})
after averaging over states in $H_D$. Such representation of the extended superoperator
clarifies physical essence of the dephasing processes as modulation of the states
$\ket i\ket i$ ¢ $H_A\otimes H_B$ by the internal states $\ket{\ket i}$, which define
an additional quantum ``phase" depending, in general, on $i$.

The states of the micro-variables of the apparatus are described in accordance with
Eq.\ (\ref{EABD}) by the partial density matrices:
\begin{equation}\label{micro}
\hat\rho_D=\sum_i p_i\ket{\ket i}\bra{\bra i},
\end{equation}

\noindent where probabilities $p_i=\bra i\hat\rho_A\ket i$ are determined only by
the density matrix of the measuring object and by the eigen basis of the measuring
physical variable $\hat A=\sum_i\lambda_i\ket i\bra i$. In case of the standard
non-coherent measurement it coincides (at a properly chosen basis set) with the
reduced density matrix of the object
$$ \hat\rho_{\rm red}=\sum_i \ket i \bra
i\hat\rho_A\ket i\bra i.
$$

\noindent In the opposite case of the completely coherent
measurement, $\ket{\ket i}\equiv\ket{\ket 0}$, we have
$\hat\rho_D=\ket{\ket0}\bra{\bra0}$ and, respectively, the microstates
entropy equals to zero.

In this connection, it is worth to note that the zero microstates entropy
does not prevent manifestation of physically essential macroscopic fluctuations
in the system. Besides a subset of physical variables for which $\ket{\ket0}$
is the eigenstate, there is an ``overwhelming majority" (this qualitative characteristic
can be readily concretized mathematically) of other variables
that results in quantum fluctuations, of perfectly macroscopic character inclusive.
It is clear, in principle, that any quantum state of a macro-object can be considered
as a pure state at the microscopic level in the frame of sufficiently complete
microscopic model, which includes all physical subsystems the object interacts
with.

The coherent information \cite{l} or preserved entanglement \cite{m,n} \[
I_c=S[\hat\rho_B]-S[\hat\rho_{AB}]
\]

\noindent corresponding to the entangling measurement can be expressed
via the entropy of the measured variables and the entropy of the dephased
micro-subsystem. To do that, we should keep in mind that due to the ideal
character of the measurement, the marginal density matrix of the measuring
variables coincides with the reduced one of the object. Also, because the
only source of decoherence is in the subsystem $D$ in transformation (\ref{EABD}),
the entropy of the joint density matrix $\hat\rho_{AB}$ coincides with the
entropy of this dephasing subsystem. As a result we get
\begin{equation}\label{coh}
I_c=S[\rho_{\rm red}]-S[\hat\rho_D].
\end{equation}

\noindent i.e., by contrast with the general case, in the form of
\emph{always positive} difference between entropy of the reduced
by the measurement state of the object and entropy of the
apparatus microstates. The latter is inevitably less than the
entropy of the resulted state of the measured object, because
otherwise it will not meet the ideal measurement requirements
according to which a macro-variable describing the result of the
measurement does not show classical fluctuations. Mathematically,
this means that $\hat k_B\equiv\hat k_A$, where $\hat k_B=\sum_i
i{\ket i}_B {\bra i}_B$, $\hat k_A=\sum_i i{\ket i}_A{\bra i}_A$,
and respective to each $\ket i$ states exhibit only quantum
uncertainty, which is due to the nonorthogonality of the
microstates $\ket{\ket i}$. Therefore, the microstates entropy
reaches the entropy $S[\hat\rho_{\rm red}]$ of the measuring
variable $\hat k_A$ (and, simultaneously, $\hat k_B$) only for the
case of ``maximally independent'', orthogonal, microstates
$\ket{\ket i}$. In this case, the coherent information (\ref{coh})
vanishes.

\section{\label{section:unitary}Unitary implementation of the extended superoperators}

\begin{figure}[h]
\begin{center}
 \epsfxsize=0.45\textwidth\epsfclipon\leavevmode\epsffile{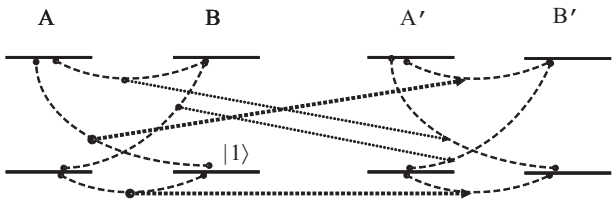}
\end{center}
\caption{\label{fig:fig1}Unitary representation ${\cal U}_C$ of the extended
cloning superoperator (\ref{EC}), which is defined on the basis of eigenstates
of two two-level systems $A$ and $B$. Basis states of the joint $A+B$ system
(dashed lines), which exist after setting the system $B$ with the
transformation ${\cal R}_0$ into the ground state $\ket1$, are transferred into
the states $\ket i\ket i$ (bold dotted arrows), whereas the rest of the states
are transferred into the states $\ket i\ket{j\ne i}$ (thin dotted arrows).}
\end{figure}

The non-reversal, invariant in respect to the apparatus' state, cloning
superoperator (\ref{C}) can be presented in the form ${\cal C}={\cal
U}_C{\cal R}_0$ as a superposition of the superoperator ${\cal R}_0={\cal
I }_A\otimes \ket1\bra1\sum_m\bra m\odot_B\ket m$, which sets an initial
state of the system $B$ into the given pure state $\ket1\bra1$, and the
respective unitary cloning superoperator ${\cal
U}_C^{}=U_C^{}\odot_{AB}U^{-1}_C$, in which the unitary transformation
$U_C$ ¢ $H_A\otimes H_B$ has the form:
\begin{equation}\label{UC}
\ket i\ket1\to\ket i \ket i,\quad \ket
i\ket{j\ne1}\to\ket{k_{ij}}\ket{l_{ij}},
\end{equation}

\noindent where two arbitrary indices $k_{ij}$, $l_{ij}$ obey the
only constraint $k_{ij}\ne l_{ij}$. This transformation is
illustrated in Fig.\ \ref{fig:fig1} on example of two two-level
systems. The extended superoperator ${\cal E}_C={\cal
C}\bigl(\odot_A\otimes\ket 1 \bra 1\bigr)$ corresponding to ${\cal
C}$ can be explicitly represented via the unitary transformation
in the bipartite system:
\begin{equation}\label{EC}
{\cal E}_C= U_C^{}\bigl(\odot_A\otimes\ket1\bra1\bigr)U^{-1}_C.
\end{equation}

\noindent From experimental point of view, it is well known that reversibility of a
physical transformation, which corresponds to the unitarity, is of great importance
for a potential implementation. This is because the reversibility is, generally,
connected with the exchange of energy and respective recoil momentum, which for the
cold atoms in traps, for instance, could lead to uncontrolled processes, up to
loosing atoms from the trap \cite{k}. However, such effects can be avoided if we
apply a non-reversal setting of the entanglement matrix ${\cal R}_0$ and,
respectively, an invariant cloning transformation ${\cal C}$ to the previously set
equilibrium state $\ket1\bra1$.

Let us now prove that the extended superoperator of entangling measurement
(\ref{EABD}) can be physically implemented with the help of unitary transformation
immediately in the system of object--ap\-pa\-ra\-tus--in\-ter\-nal variables, i.e.,
in $H_A \otimes H_B\otimes H_D$. Construction of such a transformation splits into
two steps.

First, we construct a unitary map ${\cal U}_C$ in the system
object--apparatus of the form of Eq\ (\ref{UC}) and take
into account that after this transformation in $H_A\otimes H_B$ there will
be only cloning states $\ket i\ket i$.

\begin{figure}[h]
\begin{center}
 \epsfxsize=0.45\textwidth\epsfclipon\leavevmode\epsffile{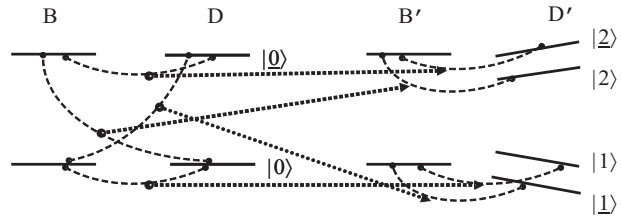}
\end{center} \caption{\label{fig:fig2}Unitary transformation $U_E$ for partial
entanglement of two qubits, where $\{\ket0,\ket{\underline0}\}$,
$\{\ket1,\ket{\underline1}\}$, and $\{\ket2,\ket{\underline2}\}$ designate
three orthogonal bases in $H_D$ arbitrary rotated to each other.}
\end{figure}

Then, after selecting an arbitrary initial state $\ket{\ket0}$ in $H_D$,
it is sufficient to construct in the subsystem $H_B\otimes H_D$ a unitary
partial entanglement operator $U_E$, which includes, in general case, dephasing
effects and fits the following relations
\begin{equation}\label{UE}
U_E\ket i\ket{\ket 0}=\ket i\ket{\ket i}
\end{equation}

\noindent for all $i=1,\dots,d$. Taking into account that vectors $\ket i$
are orthogonal to each other, such map preserves initial metric, i.e.,
orthonormalization of the transformed vectors. This guarantees that there
is a space, which maps $d^2-d$ vectors $\ket i\ket{\ket {j\ne i}}$ in a
respective arbitrary chosen basis set in a subspace orthogonal to the
$d$-mensional subspace of vectors $\ket i\ket{\ket i}$. Two-dimensional
example of such a unitary transformation is illustrated in
Fig.~\ref{fig:fig2}.

With the help of equations (\ref{UE}) and (\ref{UC}) one can easily see
that the extended superoperator of entangling measurement (\ref{EABD}) can
be written in a form of superposition of unitary transformations acting on
the object density matrices at the initial state $\ket1\ket{\ket0}$ of the
apparatus and its internal variables:
\begin{equation}\label{EABDU}
{\cal E}_M=\bigl({\cal I}_A\otimes{\cal U}_E\bigr)\bigl({\cal
U}_C\otimes{\cal
I}_D\bigr)\bigl(\odot\otimes\ket1\bra1\otimes\ket{\ket0}\bra{\bra0}\bigr).
\end{equation}

\noindent In a case of pure cloning, i.e., without any dephasing,
the unitary superoperator ${\cal U}_E$ is represented by the
identity superoperator (one should also take into account freedom
in selection the basis states, which leads to further
generalization or simplification due to the corresponding unitary
transformation of the form $\sum_j e^{i\varphi_j}\ket j\bra j$).

\section{\label{section:matrix}Matrix representation of the extended
superoperators}

From practical point of view, one of the most useful variants of the matrix
representation of the extended superoperators $A\to A{+}B$ is based on the fixed
linear basis $\hat e_k^A$ for determining the input states
$\hat\rho_A=\sum\rho_n^{}\hat e_n^A$. The corresponding representation of the
resulting density matrix $\hat\rho_{AB}=\sum\rho_n^{}{\cal E}\hat e_n^A$ is
determined then by the set of basis operators: $$\hat{\cal E}_n={\cal E}\hat
e_n^A,\quad \hat\rho_{AB}=\sum \rho_n^{}\hat{\cal E}_n.$$

\noindent Thus, the extended superoperators are represented by the
operator set $\hat{\cal E}_n$, $n=1,\dots,d^2$ in the space
$H_A\otimes H_B$. Operators $\hat{\cal E}_n$, in their turn, can
be represented by the corresponding matrices of $d^2\times
d^2$-dimensions (or by the matrices of highest dimension in case
of additionally extended space $H_B$).

One can clearly see that positivity of the extended operator ${\cal E}
\to\hat{\cal E}_n$ corresponds to the positivity of $\hat{\cal E}_n$ in a
positive basis $\hat e_n^A$.

Representation of the unitary extended superoperator in case of pure initial states
$\hat\rho_B^0=\ket1\bra1$ reduces simply to a set of orthogonal wave-functions in
$H_A\otimes H_B$. Really, for the two-indices symbolic representation ${\cal E}=\sum
U\bigl(\bra k\odot\ket l\otimes \ket1\bra1\bigr)U^{-1}$ we receive in the basis
$\ket k\bra l$ the following matrix representation: $\hat{\cal
E}_{kl}=\Psi_k^{}\Psi_l^+$, where $\Psi_k^{}=U\bigl(\ket k\ket1\bigr)$ is an
arbitrary, in general case, set of $d$ orthogonal vectors in a $d\times
d$-dimensional space. In particulary, for the considered above two-dimensional
unitary cloning transformation in accordance with the transformation (\ref{UE}) it
is represented by a pair of four-dimensional wave-functions in the right side of the
equation, which in the basis $\ket i\ket j$ are described by the rectangular matrix
$d\times d^2$ of the form:
\begin{equation}\label{C22}
\left(\Psi_k^{}\right)= \left(\begin{array}{cccc}
  1 & 0 & 0 & 0 \\
  0 & 0 & 0 & 1
\end{array}\right).
\end{equation}

\noindent Here, the coincidence of the states of the subsystems $A$ and
$B$ in bipartite states $\Psi_1^{}=\ket1\ket1$,  $\Psi_2^{}=\ket2\ket2$
provides an evidence of the clonal character of the resulting state.
Such dimension-saving symbolic representations are especially effective for
implementation of the calculations with the help of computer algebra, that
perform linear transformations with respective degenerate multidimensional
density matrices $\Psi_k^{}\Psi_l^+$ without any visible technical problems.

\section{\label{section:2D}Physical implementation of the entangling measurement
in a system of three qubits}

In this section, we analyze an explicit mathematical form of the
transformation, which can be used for a possible experimental
implementation of the specific realization of the extended
superoperator of the entangling quantum measurement described in
Sec.\ \ref{section:unitary}. A system of two two-level atoms in a
resonator could serve as a physical example for such an
experimental implementation. It can be well modelled by a
three-qubit system in which qubits $A$ and $B$ correspond to the
two-level atoms in the resonator and third qubit, $D$, describes
the states of the resonator mode of electromagnetic field, both
vacuum and one-photon.

The transformation ${\cal U}_C$ is given by Eq.\ (\ref{UC}) and we should only
specify the entangling superoperator ${\cal U}_D$, which in accordance with relation
(\ref{UE}) could be specifically defined by the map
\begin{equation}\label{UE2}
  \begin{array}{lcl}
    \ket1\ket{\ket0} & \to & \ket1\ket{\ket1}, \\
    \ket2\ket{\ket0} & \to & \ket1\ket{\ket2}, \\
    \ket1\ket{\ket{\underline0}} & \to & \ket1\ket{\ket{\underline1}}, \\
    \ket2\ket{\ket{\underline0}} & \to & \ket2\ket{\ket{\underline2}},
  \end{array}
\end{equation}

\noindent where underlining marks the vectors orthogonal to the initial
ones. The entanglement matrix in this case has all diagonal elements equal to
unit and the only off-diagonal element $R_{12}^{}=R_{21}^*=\left<\braket{1}{2}
\right>=q$, which does not equal to unit. Transformation
for the last pair of vectors can vary from shown above by an arbitrary
unitary transformation in the subspace of the respective output
pair  of the states $\ket1\ket{\ket{\underline1}}$, $\ket2\ket{\ket{\underline2}}$.

Combining transformations ${\cal U}_C$ and ${\cal U}_D$, we receive the resulting
unitary map $U_{CD}=\bigl(\hat I_A\otimes U_D\bigr)\bigl(U_C\otimes\hat I_D \bigr)$:
\begin{equation}\label{UEC}
\begin{array}{clccc}
&ABD & \Rightarrow & A'B'D' & \\
\oplus&\ket1\ket1\ket{\ket0} & \to & \ket1\ket1\ket{\ket1}, & \oplus \\
&\ket1\ket1\ket{\ket{\underline0}} & \to & \ket1\ket1\ket{\ket{\underline1}}, &\\
&\ket1\ket2\ket{\ket0} & \to & \ket1\ket2\ket{\ket{2}}, &\\
&\ket1\ket2\ket{\ket{\underline0}} & \to & \ket1\ket2\ket{\ket{\underline2}}, &\\
\oplus&\ket2\ket1\ket{\ket0} & \to & \ket2\ket2\ket{\ket2}, & \oplus \\
&\ket2\ket1\ket{\ket{\underline0}} & \to & \ket2\ket2\ket{\ket{\underline2}}, &\\
&\ket2\ket2\ket{\ket0} & \to & \ket2\ket1\ket{\ket1}, &\\
&\ket2\ket2\ket{\ket{\underline0}}&\to&\ket2\ket1\ket{\ket{\underline1}}.&
  \end{array}
\end{equation}

\noindent Symbols $\oplus$ mark here the states, which exist at
the input and are formed at the output due to the transformations
of the initial states of the form $\ket{\psi_A}\ket1 \ket0$ that
are used in our model system. As a result, only two states out of
the entire space $H_A\otimes H_B\otimes H_D$ are used both at the
input and output. It is worth to note that definition of the
transformation (\ref{UEC}) is not a unique one because in the
corresponding inactive $6d$-subspace could be defined any
arbitrary unitary transformation.

With the accuracy up to the local transformations, the unitary map
(\ref{UEC}) in orthogonal basis $\ket k\ket l \ket{\ket m}$,
$m=0,\underline0\,$ (specifically,
$\ket{\ket1}{=}\ket{\ket0}{=}(1,0)$,
$\ket{\ket2}{=}(q,\sqrt{1-|q|^2})$,
$\ket{\ket{\underline1}}{=}\ket{\ket{\underline0}}{=}(0,1)$,
$\ket{\ket{\underline2}}=(-\sqrt{1-|q|^2},q^*)$) the corresponding
matrix representation has the form:
\begin{equation}\label{UCD}
U_{CD}=\left(\begin{array}{cccccccc}
  1 & 0 & 0 & 0 & 0 & 0 & 0 & 0 \\
  0 & 1 & 0 & 0 & 0 & 0 & 0 & 0 \\
  0 & 0 & 1 & 0 & 0 & 0 & 0 & 0 \\
  0 & 0 & 0 & 1 & 0 & 0 & 0 & 0 \\
  0 & 0 & 0 & 0 & 0 & 0 & 1 & 0 \\
  0 & 0 & 0 & 0 & 0 & 0 & 0 & 1 \\
  0 & 0 & 0 & 0 & q & -\sqrt{1-|q|^2}  & 0 & 0 \\
  0 & 0 & 0 & 0 &  \sqrt{1-|q|^2} & q^* & 0 & 0
\end{array}\right).
\end{equation}

\noindent Matrix representation of the corresponding extended superoperator ${\cal
E}=U_{CD}^{}\bigl(\odot\otimes\ket1\bra1\otimes \ket0\bra0\bigr) U_{CD}^{-1}$
results, keeping in mind its unitarity and with the help of
Sec.~\ref{section:matrix}, in two 8-dimensional vectors marked by symbol $\oplus$ in
the right-side of the equation (\ref{UEC}):
$$\left(\Psi_k\right)=\left(\begin{array}{cccccccc}
  1 & 0 & 0 & 0 & 0 & 0 & 0 & 0 \\
  0 & 0 & 0 & 0 & 0 & 0 & q & \sqrt{1-|q|^2}
\end{array}\right).$$

\noindent Second vector determines a dephasing influence of the
two-level subsystem $D$ on the cloning process because the
complete state $\Psi_2=\ket2\ket2\ket{\ket2}$ has some phase
disturbance due to the difference of state $\ket{\ket2}$ of the
subsystem $D$ from $\ket{\ket1}$ in the state
$\Psi_1=\ket1\ket1\ket{\ket1}$. In general case, after the
entangling measurement we have the output, which is intermediate
between purely quantum, i.e., coherent, representation of the
output information and classical, i.e., completely dephased
representation. Module of the parameter $q$ sets the degree of
coherency, whereas its phase---freedom in choosing the phases of
the cloned states. At $q=0$ we have the standard projective
measurement.

\section{\label{section:conclusions}Conclusions}

In conclusion, we have shown that mathematical technique based on
the extended superoperators fits well for describing physical
implementations of the entangling quantum measurements, both in
case of explicitly known state of the apparatus and without any
dependence of the measurement results on its state.

It is shown that the extended superoperator of the entangling measurement has most
valuable from physical point of view information representation defined  with only a
set of state vectors in a joint three-partite system
``object--ap\-pa\-ra\-tus--in\-ter\-nal degrees of freedom of the apparatus", where
the internal degrees of freedom in $d\times d\times d$-dimensional Hilbert space
($d$ is the number of measured states) $H_A\otimes H_B\otimes H_D$ cause the
dephasing.

It is argued that the coherent information taken at the entangling measurement is
represented as ever \emph{positively} defined difference between
taken at the measurement classical information and entropy of
the internal dephasing variables.

Possible physical realization in a simple three-qubit model, which
implements the entangling quantum measurement transformation with an
arbitrary degree of entanglement is examined. Two qubits in the model
correspond to the two two-level atoms in a resonator, whereas the third
qubit models the quantum microstructure of the apparatus. The model allows
demonstration of a totally controllable transition from the completely
coherent measurement in the form of the quantum entanglement towards the
standard quantum measurement in a form of wave-function collapse. It could
also be useful in experiments studying non-reversal and decoherence
processes under maximally controllable conditions.

{\bf Acknowledgements} This work was supported in part by the Russian
Foundation for Basic Research under Grant Nos. 01--02--16311, 02--03--32200,
and by INTAS under Grant No. INFO 00--479.

\end{document}